\documentclass[english]{article}
\usepackage{amsthm}
\usepackage{amsmath}
\usepackage{units}
\usepackage{fullpage}
\usepackage{setspace}
\theoremstyle{plain}
\newtheorem{thm}{\protect\theoremname}[section]
  \theoremstyle{definition}
  \newtheorem{defn}[thm]{\protect\definitionname}
  \theoremstyle{plain}
  \newtheorem{prop}[thm]{\protect\propositionname}
  \theoremstyle{remark}
  \newtheorem{rem}[thm]{\protect\remarkname}

\providecommand{\definitionname}{Definition}
\providecommand{\propositionname}{Proposition}
\providecommand{\remarkname}{Remark}
\providecommand{\theoremname}{Theorem}

\begin{document}

\title{A Note on Almost Perfect Probabilistically Checkable Proofs of Proximity}

\author{Shlomo Jozeph}
\maketitle
\begin{abstract}
Probabilistically checkable proofs of proximity (PCPP) are proof systems
where the verifier is given a 3SAT formula, but has only oracle access
to an assignment and a proof. The verifier accepts a satisfying assignment
with a valid proof, and rejects (with high enough probability) an
assignment that is far from all satisfying assignments (for any given
proof).

In this work, we focus on the type of computation the verifier is
allowed to make. Assuming $\mathsf{P}\neq\mathsf{NP}$, there can
be no PCPP when the verifier is only allowed to answer according to
constraints from a set that forms a CSP that is solvable in \textsf{P}.
Therefore, the notion of PCPP is relaxed to almost perfect probabilistically
checkable proofs of proximity (APPCPP), where the verifier is allowed
to reject a satisfying assignment with a valid proof, with arbitrary
small probability.

We show, unconditionally, a dichotomy of sets of allowable computations:
sets that have APPCPPs (which actually follows because they have PCPPs)
and sets that do not. This dichotomy turns out to be the same as that
of the Dichotomy Theorem, which can be thought of as dividing sets
of allowable verifier computations into sets that give rise to \textsf{NP}-hard
CSPs, and sets that give rise to CSPs that are solvable in \textsf{P}.
\end{abstract}

\section{Introduction}

PCP of proximity \cite{BsGHSV04,DR06} is a proof system where the
verifier is given a 3SAT formula (or, equivalently, a circuit), and
oracle access to an assignment and a proof. The verifier is expected
to distinguish between two cases:
\begin{itemize}
\item (Completeness) The assignment satisfies the formula (we may also require
that the proof is valid), for which the verifier must always accept.
\item (Soundness) The assignment is far from all satisfying assignments,
for which the verifier must reject with a constant probability (e.g.
$\nicefrac{1}{2}$), regardless of the given proof.
\end{itemize}
Usually, there are more limitations on the verifier, such as the amount
of randomness it is allowed to use (e.g. logarithmic, polylogarithmic),
and the number of location the verifier may query (e.g. constant,
logarithmic).

In this work, we will not limit the verifier in these ways, but instead
we limit the type of calculation the verifier is allowed to make.
We fix a set of allowable computations of the verifier, who can only
choose a computation from the set. For example, if the set contains
a single function, ONE\_IN\_THREE (which outputs 1 if exactly one
of its three inputs are 1, and 0 otherwise) the verifier may query
any three locations from the assignment and the proof (based on random
coins and any computation), but the verifier accepts iff exactly one
of the locations is set to true. It can be shown that such a PCPP
exists.

In another example, which we will call the $k$-linear PCPP, the set
contains all affine functions (modulo 2) over $k$ bits. In this case,
the verifier chooses a bit $b\in\left\{ 0,1\right\} $ and up to $k$
locations to query. These choices can be made in any way. However,
the verifier must accept iff the sum of the values in the chosen locations
is equal to $b$ (modulo 2).

A $k$-linear PCPP whose verifier uses a logarithmic amount of randomness
and polynomial time cannot exist, under the assumption of $\mathsf{P}\neq\mathsf{NP}$:
Suppose otherwise. We may treat each query of the verifier as a linear
equation over the locations queried (the XOR of the locations quried
should be equal to $b$, as chosen by the verifier). Given a circuit,
we write all the linear equations corresponding to the verifier's
choices. Since the verifier uses a logarithmic amount of randomness,
there are polynomially many equations. If the circuit is satisfiable,
there is a way to completely satisfy the verifier, which means that
there is a solution for all equations. Otherwise, all assignments
are far from satisfying, so there is no way to solve all equations.
If a solution exists, it can be found via Gaussian elimination in
polynomial time. Therefore, there is an algorithm to solve the circuit
satisfaction problem in polynomial time.%
\footnote{Under the Exponential Time Hypothesis, a similar proof shows that
even $n^{\mathrm{1-\delta}}$ randomness is not enough.%
}

Hence, if we wish to consider arbitrary limitations of the power of
the verifier, we should weaken the requirements of PCPPs. Taking inspiration
from H{\aa}stad \cite{H97}, who showed that it is hard to decide
if a 3LIN instance is at least $1-\epsilon$ satisfiable or at most
$\nicefrac{1}{2}+\epsilon$ satisfiable, we weaken the completeness
requirement to almost perfect completeness. That is, for every $\epsilon>0$,
we require a verifier (that may depend on $\epsilon$) that has complete
access to a 3SAT formula and oracle access to an assignment and a
proof, to decide between the two cases:
\begin{itemize}
\item The assignment satisfies the formula (and the proof is valid), for
which the verifier accepts with probability at least $1-\epsilon$.
\item The assignment is far from all satisfying assignments, for which the
verifier accepts with probability at most $\nicefrac{1}{2}$ (we even
allow at most $1-\mathrm{\omega}\left(\epsilon\right)$).
\end{itemize}
Our result shows that, unconditionally, some sets of allowable computations
do not have Almost Perfect Probabilistically Checkable Proofs of Proximity.
These sets are exactly the ones that give rise to constraint satisfaction
problems that are in $\mathsf{P}$, under the assumption $\mathsf{P}\neq\mathsf{NP}$.
On the other hand, any set that gives rise to CSP that is $\mathsf{NP}$-complete
has PCPP, which follows almost immediately from the existence of PCP
of proximity \cite{BsGHSV04,DR06} and gadget reductions from 3SAT
\cite{S78} (these reductions are stated more explicitly in \cite{CKS01}).

\subsection{Definitions}

The definitions related to constraints are based on the definitions
from \cite{CKS01}.
\begin{defn}
A \emph{constraint} is a function $f:\left\{ 0,1\right\} ^{k}\to\left\{ 0,1\right\} $.
$k$ is the \emph{arity} of the constraint. The constraint is \emph{satisfied}
by an assignment $x$ if $f\left(x\right)=1$.

We define the constraints ID and NOT which have arity 1 and return
the input and its negation, respectively.
\end{defn}

\begin{defn}
A \emph{constraint set} is a non-empty set of constraints.
\end{defn}
For example, the constraint set corresponding to 2LIN contains two
constraints, $f_{1}$ and $f_{2}$, with $f_{1}\left(x,y\right)=x\oplus y$
and $f_{2}\left(x,y\right)=x\oplus y\oplus1$.

By the definition used in this paper, a constraint satisfaction problem
is specified by a constraint set. Another way to specify a constraint
satisfaction problem is by using a single function and allowing the
usage of all possible negations. For example, our definition may need
a constraint set containing 8 constraints to define 3SAT. It can also
be defined by a single function, $f\left(x,y,z\right)=x\vee y\vee z$,
if negations are allowed implicitly.
\begin{defn}
A \emph{constraint application} is an ordered set $\left\langle f,i_{1},\ldots,i_{k}\right\rangle $,
where $f$ is a constraint of arity $k$, and each $i_{j}$ is a natural
number indicating the name of the variable.
\end{defn}
The same variable name may appear several times in a constraint application.
\begin{defn}
Given a constraint set $S$, a \emph{formula over $S$ with $n$ variables}
is a (multi)set $P$ containing constraint applications. For each
$c\in P$, $c=\left\langle f,i_{1},\ldots,i_{k}\right\rangle $, where
$k$ is the arity of $f\in S$, and $i\in\left[n\right]$. Usually,
$S$ and $n$ will be implied from the context.

An \emph{assignment} $x\in\left\{ 0,1\right\} ^{n}$ satisfies a constraint
application $\left\langle f,i_{1},\ldots,i_{k}\right\rangle $ if
$f\left(x_{i_{1}},\ldots\text{,}x_{i_{k}}\right)=1$. A formula is
$\alpha$-satisfied by an assignment $x$ if an $\alpha$-fraction
of the constraints in the formula are satisfied. A formula is $\alpha$-satisfiable
if there is an assignment that $\alpha$-satisfies it. In the case
of $\alpha=1$ we may omit $\alpha$.
\end{defn}
A formula may contain several copies of the same constraint application.
As a shorthand, we may say we give a constraint weight $t$ when we
mean that there are $t$ copies of it in the formula. This can be
generalized to non integer weights, since we will only care about
fixed precision of weights (so we multiply all weights by a constant,
and round to the nearest integer).
\begin{defn}
Given a constraint set $S$ and $0\leq\varsigma<\kappa\leq1$, an
\emph{$\left(S,\kappa,\varsigma\right)$-Constraint Satisfaction Problem}
(\emph{$\left(S,\kappa,\varsigma\right)$}-CSP) is the following problem:
Given a formula over $S$, decide whether there is an assignment to
the variables satisfying at least $\kappa$-fraction of the constraint
applications, or any assignment satisfies at most $\varsigma$-fraction
of the variables.

A constraint set $S$ will be called $\mathsf{NP}$\emph{-hard} if
, for some $\varsigma$, solving the \emph{$\left(S,1,\varsigma\right)$}-CSP
is $\mathsf{NP}$-hard. A constraint set $S$ will be called \emph{$\mathsf{APX}$-hard}
if solving the \emph{$\left(S,\kappa,\varsigma\right)$}-CSP, for
some constants $0<\varsigma<\kappa<1$, is $\mathsf{NP}$-hard.
\end{defn}
Note that Every $\mathsf{NP}$-hard constraint set is an $\mathsf{APX}$-hard
constraint set but the other direction is not true.
\begin{defn}
Given a constraint set $S$, an \emph{$S$-verifier} $V$ is a function
on a 3SAT formula $\varphi$, and a random string $r$ (of length
bounded by some function of $\varphi$). The verifier outputs $f^{r}\in S$
of arity $k$ and indices $i_{1}^{r},\dots,i_{k}^{r}$. For a string
$s$, the \emph{acceptance probability} of $V(\varphi)$ (the probability
is on the random strings $r$) on $s$ is the probability (over $r$)
that $f^{r}\left(s_{i_{1}^{r}},\dots,s_{i_{k}^{r}}\right)=1$.
\end{defn}
Note that an $S$-verifier defines a formula over $S$, by considering
each index the verifier outputs as a variable, and defining a constraint
application for every random string $r$ by the output of the verifier
on $r$. Then, the\emph{ }acceptance probability of $V(\varphi)$
on $s$ is the fraction of constraints satisfied by $s$.
\begin{defn}
An \emph{$\left(S,\delta,d\right)$-APPCPP (Almost Perfect Probabilistically
Checkable Proofs of Proximity)} for $d>1$, $\delta>0$ is a set of
$S$-verifiers. For every $\epsilon>0$ small enough (that is, for
some $\Lambda>0$, for every $\Lambda>\epsilon>0$) the set contains
a verifier $V_{\epsilon}$. For every integer $n>0$, and $\varphi$,
a 3SAT formula $\varphi$ with $n$ variables, the following holds:
\begin{itemize}
\item (Completeness) If $\bar{a}\in\left\{ 0,1\right\} ^{n}$ is a satisfying
assignment to $\varphi$, then there is a $k>0$ and $\bar{\pi}\in\left\{ 0,1\right\} ^{p}$
such that the acceptance probability of $V_{\epsilon}\left(\varphi\right)$
on $\left(\bar{a},\bar{\pi}\right)$ is at least $1-\epsilon$. 
\item (Soundness) If $\bar{a}\in\left\{ 0,1\right\} ^{n}$ is $\left(1-\delta\right)$-far
from a satisfying assignment to $\varphi$, then for any $\bar{\pi}\in\left\{ 0,1\right\} ^{p}$,
the acceptance probability of $V_{\epsilon}\left(\varphi\right)$
on $\left(\bar{a},\bar{\pi}\right)$ is at most $1-d\epsilon$.
\end{itemize}
$\bar{\pi}$ is called a proof for $\bar{a}$. It is accepted in the
completeness case, and rejected in the soundness case.
\end{defn}
The definition does not restrict the verifiers in any way. For example,
the verifiers may use an exponential amount of randomness, may require
$p$ to be exponential, and the may even be uncomputable.
\begin{defn}
A constraint $f$ is said to be 
\begin{itemize}
\item \emph{0-valid} if $f\left(0,\cdots,0\right)=1$.
\item \emph{1-valid} if $f\left(1,\cdots,1\right)=1$.
\item \emph{Weakly positive} if $f$ is equivalent to a conjunction of CNF
clauses with at most one negated variable in each clause.
\item \emph{Weakly negative} if $f$ is equivalent to a conjunction of CNF
clauses with at most one non-negated variable in each clause.
\item \emph{Linear} if $f$ is equivalent to a conjunction of linear equations.
\item \emph{2CNF} if $f$ is equivalent to a conjunction of 2CNF clauses.
\item \emph{C-closed} if $f\left(x\right)=f\left(\bar{x}\right)$ for all
$x$, where $\bar{x}$ is the complement of the assignment $x$.
\end{itemize}
We use the same terminology for constraint sets, if all of the contained
constraints satisfy the respective condition.
\end{defn}

\subsection{Results}

As a first step, we state that any non-C-closed constraint set, $S$,
that is $\mathsf{NP}$-hard under the assumption \textsf{P}$\neq$\textsf{NP},
has $\left(S,\delta,d\right)$-APPCPP for some $\delta$ and any $d$.
This follows from the existence of PCPPs which have perfect completeness
and constant soundness.
\begin{prop}
\label{prop:PCPPExist}For every constraint set $S$ that is NP-hard
and not C-closed, there is a PCPP with perfect completeness and constant
soundness that only uses constraints from $S$.

For every constraint set $S$ that is $\mathsf{APX}$-hard and not
C-closed, there is a PCPP without perfect completeness (but constant
completeness and soundness) that only uses constraints from $S$.
\end{prop}
Now we state that for any non-C-closed constraint set, $S$, that
is not $\mathsf{NP}$-hard under the assumption \textsf{P}$\neq$\textsf{NP},
there is no $\left(S,\delta,d\right)$-APPCPP for any $\delta$ (and
some $d$ depending on $\delta$).
\begin{thm}
\label{thm:NoAPPCPP}The set of non-C-closed constraint set that do
not have $\left(S,\delta,d\right)$-APPCPP with some $\delta>0$,
$d>\left(\left\lceil \delta^{-1}\right\rceil +2\right)^{2}$ is the
set of non-C-closed constraint sets that are \textsf{NP}-complete,
under the assumption \textsf{P}$\neq$\textsf{NP}.
\end{thm}
Combining these two results together, we get an unconditional dichotomy
of non-C-closed constraint sets, into constraint sets that have PCPPs,
and constraint sets that do not even have APPCPPs. The results may
be extended to C-closed constraint sets, either by modifying the constraint
sets to be non-C-closed, or by modifying the definition of APPCPP
(see remark \ref{rem:C-closed}).

This dichotomy may help to explain why PCPPs are harder to construct
than PCPs. Specifically, if we assume that $\mathsf{P}=\mathsf{NP}$,
we may build any PCP using any constraint set. However, even such
an incredible assumption does not help to construct even APPCPPs for
some constraint sets.

Additionally, this dichotomy shows that PCPPs are stronger than locally
testable codes. Locally testable codes can be constructed from good
codes and PCPPs (see \cite{BsGHSV04}), but even the existence of
a locally testable and decodable code, where the tests are linear
(the Hadamard code) does not help to generate APPCPPs over linear
constraint sets (or a $k$-linear PCPP).

\subsection{Related Work}

The Dichotomy Theorem \cite{S78} states that every Boolean constraint
satisfaction problem (CSP) is either \textsf{NP}-complete or solvable
in polynomial time. Naturally, such a dichotomy is only meaningful
if $\mathsf{P}\neq\mathsf{NP}$. However, Schaefer's proof of the
Dichotomy Theorem can be thought of as showing an unconditional result,
which is a weaker notion of PCPs of proximity: Suppose we are given
a 3SAT formula (or, equivalently, a circuit), and oracle access to
an assignment and a proof. We wish to use a verifier that queries
the assignment and the proof at a few places to differentiate between
two cases:
\begin{itemize}
\item (Completeness) The assignment satisfies the formula (we may also require
that the proof is valid), for which the verifier must always accept.
\item (Soundness) The assignment does not satisfy the formula (for any given
proof), for which the verifier must reject at least once.
\end{itemize}
Then, only \textsf{NP}-hard sets may have such a weak PCPP.

Creignou \cite{C95} showed that CSPs can be classified into three
classes by approximation hardness: CSPs that are $\mathsf{NP}$-hard,
CSPs that are $\mathsf{APX}$-hard, and CSPs that can be maximized
in polynomial time. Bulatov proved the Dichotomy Theorem for CSPs
over variables that have three possible values \cite{B02}. Whether
such a dichotomy holds for other non-Boolean CSPs is open (see \cite{B11}
for a survey).

Probabilistically checkable proofs of proximity were introduced by
Ben-Sasson et al. \cite{BsGHSV04} and as assignment testers by Dinur
and Reingold \cite{DR06} as a tool helping composition of PCPs. PCPs
of proximity are required to accept any correct proof with probability
1, unlike APPCPPs. The rejection probability may be defined in various
ways. The most general one is a function of the distance of the assignment
from the closest satisfying assignment. A simpler definition requires
only a constant rejection probability if an assignment is constantly
far from satisfying assignments.%
\footnote{The proofs in this paper can be slightly simplified, if we only ask
for a constant rejection probability.%
}

Guruswami proved a theorem (Theorem 1 in \cite{G06}) similar to proposition
\ref{prop:no2cnf}, with looser parameters.

\section{Proofs}
\begin{proof}
[Proof sketch of proposition \ref{prop:PCPPExist}]PCPPs with perfect
completeness and constant soundness where the verifier, $V$, makes
a constant number of queries exist (see, for example \cite{BsGHSV04}).
There are gadget reductions from 3SAT to each $\mathsf{NP}$-hard
constraint set $S$ (which is not C-closed), as shown by \cite{S78}.

Consider a query of the verifier, $V$ that queries $k$ location.
WLOG, we may assume that the verifier queries makes $<2^{k}$ $k$SAT
calculaions. Using a reduction from $k$SAT to 3SAT (e.g. adding $k-3$
variables) and the gadget reuction from 3SAT to $S$, we change the
this query of $V$ to use only constraints from $S$. This costs us
a polynomial (in $k$) amount of calculation to be done by the verifier
per (which we ignore in this paper) and a constant (depending on $k$)
addition of variables to the proof, per each query.

Similarly, for an $\mathsf{APX}$-hard constraint set $S$, a PCPP
with perfect completeness and constant soundness can be transformed
into a PCPP without perfect completeness (but constant completeness
and soundness) over $S$ using a gadget reduction from 3SAT to $S$
which was shown to exist by \cite{CKS01}.
\end{proof}

\begin{proof}
[Proof of Theorem \ref{thm:NoAPPCPP}]By Schaefer's classification
\cite{S78}, all constraint sets that are not NP-hard, under the assumption
\textsf{P}$\neq$\textsf{NP}, are constraint sets of the following
six types: Linear, Weakly positive, Weakly Negative, 2CNF, 1-valid,
and 0-valid.

Linear constraint sets do not have APPCPPs due to proposition \ref{prop:nolinear}.
Weakly positive and negative constraint sets do not have APPCPPs due
to proposition \ref{prop:nohorn}. 2CNF constraint sets do not have
APPCPPs due to proposition \ref{prop:no2cnf}. 1-valid constraint
sets accepts the all one assignment, so there's no $\left(C,\delta,d\right)$-APPCPP
for $C$ that is 1-valid (since there are 3SAT formulas that only
accept the all zero assignment). Similarly, there is no $\left(C,\delta,d\right)$-APPCPP
for $C$ that is 0-valid.
\end{proof}
Note that the classification of C-closed constraint sets into those
the have APPCPPs and those who do not does not depend on whether $\mathsf{P=NP}$
or $\mathsf{P\neq NP}$.
\begin{rem}
\label{rem:C-closed}If a constraint set $S$ is C-closed, by adding
any constraint $c$ that is not C-closed, $S\cup\left\{ c\right\} $
becomes non-C-closed. The constraints ID and NOT are linear, weakly
positive, weakly negative and 2CNF. The constraint ID is 1-valid,
the constraint NOT is 0-valid. Therefore, to handle the case of C-closed
constraint sets, we may allow the addition of the constraint ID or
the constraint NOT without affecting the NP-hardness of the constraint
set, thus converting a C-closed constraint set into a non-C-closed
constraint set.

Another method to deal with C-closed constraint set is to relax the
notion of APPCPP. That is, to accept assignments that are satisfying
or the bit-wise inversion of satisfying assignment, and reject assignments
that are far from both satisfying assignments and bit-wise inversions
of satisfying assignments. This requires a further relaxation of APPCPP
to only reject assignments that are $\nicefrac{1}{2}-\delta$ far,
since if there is a satisfying assignment, every string is $\nicefrac{1}{2}$-close
to it or its bit-wise inversion.

The proofs can be adapted to this model by defining a 3SAT formula
on twice the number of variables: on one half of the variables the
constraints are the same and the second half of the variables need
to be true (by using the clause $x$ for every variable $x$). Then,
the verifier will accept an all true assignment with some proof (for
the case of weakly negative and 0-valid, the second half need to be
false).
\end{rem}

\subsection{Linear Constraint Sets}
\begin{prop}
\label{prop:nolinear}There are no\emph{$\left(C,\delta,\left\lceil \delta^{-1}\right\rceil +2\right)$}-APPCPP,
for $C$ that is equivalent to a conjunction of linear equations. \end{prop}
\begin{proof}
Let $\varphi$ be a 2CNF formula on $mn$ variables, $\left\{ x_{i}^{j}\right\} $,
where $1\leq i\leq n$ and $1\leq j\leq m$, for an odd $m$ such
that $\left\lceil \delta^{-1}\right\rceil \leq m\leq\left\lceil \delta^{-1}\right\rceil +1$.
For every $x_{i}^{j}$ and $x_{i'}^{j}$, $\varphi$ contains constraints
requiring that $x_{i}^{j}=x_{i'}^{j}$ ($\overline{x_{i}^{j}}\vee x_{i'}^{j}$
and $x_{i}^{j}\vee\overline{x_{i'}^{j}}$). For every $j\neq j'$
$\varphi$ contains a constraint requiring that at most one of $x_{i}^{j}$
and $x_{i'}^{j'}$ is true ($\overline{x_{i}^{j}}\vee\overline{x_{i'}^{j'}}$).
The only assignments completely satisfying $\varphi$ are the assignments$\left\{ \alpha_{j}\right\} _{j=0}^{m}$,
where $\alpha_{j}\left(x_{i}^{j}\right)$ is true and for any $j'\neq j$,
$\alpha_{j}\left(x_{i}^{j'}\right)$ is false (the assignment $\alpha_{0}$
is always false).

Suppose that there is a \emph{$\left(C,\delta,\left\lceil \delta^{-1}\right\rceil +2\right)$}-APPCPP
for some constraint set $C$ that is equivalent to a conjunction of
linear equations. Then, there is some $\Lambda>0$, such that for
any $\Lambda>\epsilon>0$ there is a formula $\psi_{\epsilon}$ (defined
by the verifier $V_{\epsilon}$) over $C$ with the variables $\left\{ x_{i}^{j}\right\} $
and additional auxiliary variables, such that for every $\alpha_{j}$
there is a proof $\pi_{j}^{\epsilon}$ (on the auxiliary variables)
such that $\left(\alpha_{j},\pi_{j}^{\epsilon}\right)$ $\left(1-\epsilon\right)$-satisfies
$\psi_{\epsilon}$, and for any $\alpha$ which is $\delta$-far from
all $\alpha_{j}$ and any $\pi$, $\left(\alpha,\pi\right)$ at most
$\left(1-\left(\left\lceil \delta^{-1}\right\rceil +2\right)\epsilon\right)$-satisfies
$\psi_{\epsilon}$. We show that for the assignment $\beta$ that
gives every variable from $\left\{ x_{i}^{j}\right\} $ true, and
for some $\pi$, specifically for $\pi=\oplus\pi_{j}^{\epsilon}$,
at least $1-\left(\left\lceil \delta^{-1}\right\rceil +1\right)\epsilon$
of the constraints are true in $\psi\left(\beta,\pi\right)$. Note
that $\beta=\oplus\alpha_{j}$.

The first step is removing all constraints from $\psi_{\epsilon}$
that are not satisfied when assigning the variables $\left(\alpha_{j},\pi_{j}^{\epsilon}\right)$,
for all $j>0$. This removes at most $m\epsilon$ constraints from
$\psi$. We show that all of the other constraints (an $1-m\epsilon$
fraction) are satisfied when using the assignment $\left(\beta,\pi\right)$.
Consider a linear constraint $\oplus y_{k}=b$, where $\left\{ y_{k}\right\} $
are a set of variables and $b\in\left\{ 0,1\right\} $. An assignment
$\alpha:\left\{ y_{k}\right\} \to\left\{ 0,1\right\} $ satisfying
this constraint satisfies $\oplus_{k}\alpha\left(y_{k}\right)=b$.
Given $m$ assignments satisfying this constraint $\gamma_{j}:\left\{ y_{k}\right\} \to\left\{ 0,1\right\} $,
the assignment $\gamma=\oplus\gamma_{j}$ also satisfies the constraint:
\[
\oplus_{k}\gamma\left(y_{k}\right)=\oplus_{k}\left(\oplus_{j}\gamma_{j}\left(y_{k}\right)\right)=\oplus_{j}\left(\oplus_{k}\gamma_{j}\left(y_{k}\right)\right)=\oplus_{j}b=b
\]
since $m$ is odd. A conjunction of linear constraints satisfied by
all the $\gamma_{j}$'s is also satisfied by $\gamma$.

Therefore, the assignment $\left(\beta,\pi\right)=\left(\oplus\alpha_{j},\oplus\pi_{j}^{\epsilon}\right)$
satisfies all of the constraints in $\psi$ not removed in the first
step. However, $\beta$ is $1-\nicefrac{1}{m}$ far from any of the
$\alpha_{j}$'s, contradicting the definition of a \emph{$\left(C,\delta,\left\lceil \delta^{-1}\right\rceil +2\right)$}-APPCPP.
\end{proof}

\subsection{Weakly Positive and Weakly Negative Constraint Sets}
\begin{prop}
\label{prop:nohorn}There are no\emph{$\left(C,\delta,\left\lceil \delta^{-1}\right\rceil +1\right)$}-APPCPP,
for $C$ that is equivalent to a conjunction of Horn clauses. \end{prop}
\begin{proof}
Let $\varphi$ be a 2CNF formula on $mn$ variables, $\left\{ x_{i}^{j}\right\} $,
where $1\leq i\leq n$ and $1\leq j\leq m$, for $m=\left\lceil \delta^{-1}\right\rceil $.
For any $x_{i}^{j}$, $x_{i'}^{j}$ $\varphi$ contains constraints
requiring that $x_{i}^{j}=x_{i'}^{j}$ ($\overline{x_{i}^{j}}\vee x_{i'}^{j}$
and $x_{i}^{j}\vee\overline{x_{i'}^{j}}$). For every $j\neq j'$
$\varphi$ contains a constraint requiring that at most one of $x_{i}^{j}$
and $x_{i'}^{j'}$ is true ($\overline{x_{i}^{j}}\vee\overline{x_{i'}^{j'}}$).
The only assignments completely satisfying $\varphi$ are the assignments$\left\{ \alpha_{j}\right\} _{j=0}^{m}$,
where $\alpha_{j}\left(x_{i}^{j}\right)$ is true and for any $j'\neq j$
$\alpha_{j}\left(x_{i}^{j'}\right)$ is false (the assignment $\alpha_{0}$
is always false).

Suppose that there is a \emph{$\left(C,\delta,\left\lceil \delta^{-1}\right\rceil +1\right)$}-APPCPP
for some constraint set $C$ that is equivalent to a conjunction of
CNF clauses, where each clause has at most one negated variable. Then,
there is some $\Lambda>0$, such that for any $\Lambda>\epsilon>0$
there is a formula $\psi_{\epsilon}$ (defined by the verifier $V_{\epsilon}$)
over $C$ with the variables $\left\{ x_{i}^{j}\right\} $ and additional
auxiliary variables, such that for every $\alpha_{j}$ there is a
proof $\pi_{j}^{\epsilon}$ (on the auxiliary variables) such that
$\left(\alpha_{j},\pi_{j}^{\epsilon}\right)$ $\left(1-\epsilon\right)$-satisfies
$\psi_{\epsilon}$ and for any $\alpha$ which is $\delta$-far from
all $\alpha_{j}$ and any $\pi$, $\left(\alpha,\pi\right)$ at most
$\left(1-\left(\left\lceil \delta^{-1}\right\rceil +1\right)\epsilon\right)$-satisfies
$\psi_{\epsilon}$. We show that for the assignment $\beta$ that
gives every variable from $\left\{ x_{i}^{j}\right\} $ true, and
for some $\pi$, specifically for $\pi=\vee\pi_{j}^{\epsilon}$, at
least $1-\left\lceil \delta^{-1}\right\rceil \epsilon$ of the constraints
are true in $\psi\left(\beta,\pi\right)$. Note that $\beta=\vee\alpha_{j}$.

The first step is removing all constraints from $\psi_{\epsilon}$
that are are not satisfied when assigning the variables $\left(\alpha_{j},\pi_{j}^{\epsilon}\right)$,
for all $j>0$. This removes at most $m\epsilon$ fraction of the
constraints from $\psi$. We show that all of the other constraints
(an $1-m\epsilon$ fraction) are satisfied when using the assignment
$\left(\beta,\pi\right)$. Consider a clause $\vee_{k>0}y_{k}\vee\overline{y_{0}}$,
where $\left\{ y_{k}\right\} $ are a set of variables. Given $m$
assignments satisfying this clause $\gamma_{j}:\left\{ y_{k}\right\} \to\left\{ 0,1\right\} $,
the assignment $\gamma=\vee\gamma_{k}$ also satisfies the clause:
If for all $k$ $\gamma_{k}\left(y_{0}\right)=0$, then $\gamma\left(y_{0}\right)=0$,
so $\gamma$ satisfies the clause. Otherwise, for some $k$ there
is an $i$ such that $\gamma_{k}\left(y_{i}\right)=1$, so $\gamma\left(y_{i}\right)=1$,
and $\gamma$ satisfies the clause. The second argument also works
for a clause of the form $\vee_{k}y_{k}$. A conjunction of Horn clauses
satisfied by all the $\gamma_{j}$'s is also satisfied by $\gamma$.

Therefore, the assignment $\left(\beta,\pi\right)=\left(\vee\alpha_{j},\vee\pi_{j}^{\epsilon}\right)$
satisfies all of the constraints in $\psi$ not removed in the first
step. However, $\beta$ is $1-\nicefrac{1}{m}$ far from any of the
$\alpha_{j}$'s, contradicting the definition of a \emph{$\left(C,\delta,\left\lceil \delta^{-1}\right\rceil +1\right)$}-APPCPP.

Similarly, by negating all assignments used and all variables in the
formulas, it can be shown that there is no \emph{$\left(C,1-\delta,\left\lceil \delta^{-1}\right\rceil +1\right)$}-APPCPP,
when $C$ is equivalent to a conjunction clauses which have at most
one non-negated variable.
\end{proof}

\subsection{2CNF Constraint Sets}
\begin{prop}
\label{prop:no2cnf}There are no\emph{$\left(C,\delta,\left(\left\lceil \delta^{-1}\right\rceil +2\right)^{2}\right)$}-APPCPP,
for $C$ that is equivalent to a conjunction 2CNF clauses.\end{prop}
\begin{proof}
Let $\varphi$ be a 2CNF formula on $mn$ variables, $\left\{ x_{i}^{j}\right\} $,
where $1\leq i\leq n$ and $1\leq j\leq m$, for an odd $m$ such
that $\delta^{-1}\leq m\leq\left\lceil \delta^{-1}\right\rceil +1$
. For any $x_{i}^{j}$, $x_{i'}^{j}$, $\varphi$ contains constraints
requiring that $x_{i}^{j}=x_{i'}^{j}$ ($\overline{x_{i}^{j}}\vee x_{i'}^{j}$
and $x_{i}^{j}\vee\overline{x_{i'}^{j}}$) and for every distinct
$j,j',j''$, $\varphi$ contains a constraint requiring that at most
two of $x_{i}^{j},x_{i'}^{j'},x_{i''}^{j''}$ are true ($\overline{x_{i}^{j}}\vee\overline{x_{i'}^{j'}}\vee\overline{x_{i''}^{j''}}$).
The only assignments completely satisfying $\varphi$ are the assignments$\left\{ \alpha_{j,k}\right\} $,
where $\alpha_{j,k}\left(x_{i}^{t}\right)$ is true iff $t\in\left\{ j,k\right\} $
(and the assignment $\alpha_{0}$ which gives 0 to all variables).
Note that the assignment $\alpha_{j,k}$ and $\alpha_{k,,j}$ are
the same, and we assume that $\pi_{j,k}^{\epsilon}$ and $\pi_{k,,j}^{\epsilon}$
are the same as well.

Suppose that there is a \emph{$\left(C,\delta,\left(\left\lceil \delta^{-1}\right\rceil +2\right)^{2}\right)$}-APPCPP
for some constraint set $C$ that is equivalent to a conjunction of
2CNF clauses. Then, there is some $\Lambda>0$, such that for any
$\Lambda>\epsilon>0$ there is a formula $\psi_{\epsilon}$ (defined
by the verifier $V_{\epsilon}$) over $C$ with the variables $\left\{ x_{i}^{j}\right\} $
and additional auxiliary variables, such that for every $\alpha_{j,k}$
there is a proof $\pi_{j,k}^{\epsilon}$ (on the auxiliary variables)
such that $1-\epsilon$ of the constraints are true in $\psi\left(\alpha_{j,k},\pi_{j,k}^{\epsilon}\right)$.
We show that for the assignment $\beta$ that assigns every variable
from $\left\{ x_{i}^{j}\right\} $ true, and for some $\pi$, at least
$1-\left(\left\lceil \delta^{-1}\right\rceil +2\right)^{2}\epsilon$
of the constraints in $\psi_{\epsilon}$ are satisfied by $\left(\beta,\pi\right)$.

The first step is removing all constraints from $\psi$ that are are
not satisfied when assigning the variables $\left(\alpha_{j,k},\pi_{j,k}\right)$
for all $j,k$. This removes at most $m\left(m+1\right)\epsilon/2$
fraction of constraints from $\psi$. We show that all of the other
constraints (an $1-m\left(m+1\right)\epsilon/2$ fraction) can be
satisfied with $\left(\beta,\pi\right)$, for some $\pi$ that will
be constructed.

Any 2CNF formula can be represented by a graph on the literals with
two directed edges $\left(\bar{x},y\right)$ and $\left(\bar{y},x\right)$
for every clause $x\vee y$, where each directed edge mean implication,
that is, if the first literal is true, the second must be true as
well (and if the first literal is false, the implication is true).
Additionally, a variable and its negation must have opposite values.

Let $\mathcal{M}_{m}$ be the set of all $m\times m$ 0/1 symmetric
matrices. For $A,B\in\mathcal{M}_{m}$, we say that $A\leq B$ iff
$A_{p,q}\leq B_{p,q}$ for all $p,q$.

Given a variable $z$ of $\psi$, let $V\left(z\right)$ be the (symmetric)
matrix containing the $m^{2}$ values $z$ is assigned by $\left\{ \left(\alpha_{j,k},\pi_{j,k}\right)\right\} $.
For $A\in\mathcal{M}_{m}$, let $V_{A}$ be the subset of variables
of $\psi$ such that $V\left(z\right)=A$. Let $D^{j}$ be the matrix
such that $D_{p,q}^{j}=1$ iff $j\in\left\{ p,q\right\} $. Using
this notation, we state the following facts:
\begin{itemize}
\item If $A\ne B$, $V_{A}\cap V_{B}=\emptyset$ (since a variable cannot
be assigned two different values).
\item For all $i,j$, $V\left(x_{i}^{j}\right)=D^{j}$, so $x_{i}^{j}\in V_{D^{j}}$
(by definition of $\alpha_{j,k}$).
\item If $x\in V_{A}$ then $\overline{x}\in V_{B}$, for $B$ such that
$A+B$ is the all ones matrix (since negations of literals must have
opposite values).
\end{itemize}
Since we removed all constraints that are not satisfied in one of
the $\left(\alpha_{j,k},\pi_{j,k}\right)$, if there is an index $\left(q,p\right)$
such that $A_{q,p}>B_{q,p}$ there cannot be an edge from $V_{A}$
to $V_{B}$ (otherwise,$A_{q,p}=1$, $B_{q,p}=0$, so $\left(\alpha_{q,p},\pi_{q,p}\right)$
assigns true to the variables in $V_{A}$ but false to the variables
in $V_{B}$ which does not satisfy the implication constraint of the
edge). Thus, there can only be edges from $V_{A}$ to $V_{B}$ if
$A\leq B$.

Now we can construct $\pi$. For each $A\in\mathcal{M}_{m}$,$\pi$
assigns the same value to all variables in $V_{A}$. For all $j$,
$\pi$ assigns 1 to the variables in $V_{D^{j}}$. To ensure that
all on a directed path from $V_{D^{j}}$are satisfied, For all $A$
and $j$ such that $D^{j}\leq A$, $\pi$ assigns 1 to the variables
in $V_{A}$. Since $\pi$ must be a valid assignment, $\pi$ assigns
0 to the variables in $V_{1-D^{j}}$ and all variables in $A$ such
that $A\leq1-D^{j}$. Note that there are no $A,j$ such that $D^{j}\leq A\leq1-D^{j}$,
since then $D^{j}\leq1-D^{j}$ and $2D^{j}\leq1$, which is false
($D^{j}$ has some elements of value 1). For all other variables,
$\pi$ gives the variables in $V_{A}$ the majority of entries in
$A$. By the construction of $\pi$, for all $A\leq B$, an edge from
$A$ to $B$ must be satisfied by $\pi$, and for all $A+B=1$, the
variables in $V_{A}$ have the opposite assignment to the variables
in $V_{B}$ (either by the majority condition, or by the fact that
if $D^{j}\leq A$ then $B\leq1-D^{j}$). The variables of $\beta$
are contained in $\cup D_{j}$, and are set to 1, so all the edges
of the graph are satisfied (a $1-m\left(m+1\right)\epsilon/2$ fraction
of the original constraints), but $\beta$ is $1-\nicefrac{2}{m}$
far from any satisfying assignment to $\psi$.
\end{proof}

\subsection*{Acknowledgments}

Work supported in part by the Israel Science Foundation (grant No.
621/12).

\bibliographystyle{plain}
\bibliography{u}

\end{document}